\documentclass[a4paper,twocolumn, secnumarabic, aps, superscriptaddress,
amssymb, amsmath, nobibnotes, showpacs]{revtex4}
\usepackage{graphicx}
\usepackage{dcolumn}

\begin{document}
\title{Facilitated diffusion of DNA-binding proteins: Simulation
of large systems}

\author{Holger Merlitz}
\email{merlitz@gmx.de}
\affiliation{Softmatter Lab, Department of Physics, Xiamen University,
Xiamen 361005, P.R.\ China}
\author{Konstantin V.\ Klenin}
\affiliation{Division of Biophysics of Macromolecules,
German Cancer Research Center, D-69120 Heidelberg, Germany}
\author{Chen-Xu Wu}
\email{cxwu@jingxian.xmu.edu.cn}
\affiliation{Softmatter Lab, Department of Physics, Xiamen University,
Xiamen 361005, P.R.\ China}
\author{J\"org Langowski}
\affiliation{Division of Biophysics of Macromolecules,
German Cancer Research Center, D-69120 Heidelberg, Germany}

\date{\today}

\begin{abstract}
The recently introduced method of excess collisions (MEC) 
is modified to estimate diffusion-controlled reaction times 
inside systems of arbitrary size. The resulting MEC-E equations 
contain a set of empirical parameters, which
have to be calibrated in numerical simulations inside
a test system of moderate size. Once this is done, reaction times
of systems of arbitrary dimensions are derived by extrapolation,
with an accuracy of 10 to 15 percent. The achieved speed up, 
when compared to explicit simulations of the reaction process, 
is increasing proportional to the extrapolated volume of 
the cell. 
\end{abstract}


\maketitle

\section{Introduction} 
Diffusion controlled bio-chemical reactions play a 
central role in keeping any organism alive~\cite{riggs70,richter74}:
The transport of molecules through cell membranes, the passage of
ions across the synaptic gap, or the search carried out by drugs on the way to
their protein receptors are predominantly diffusive
processes. Further more, essentially all of the biological
functions of DNA are performed by proteins that interact
with specific DNA sequences~\cite{berg85, ptashne01},
and these reactions are diffusion-controlled.

However, it has been realized that some proteins
are able to find their specific binding sites on DNA
much more rapidly than is `allowed' by the diffusion
limit~\cite{riggs70, berg81}. It is therefore generally
accepted that some kind of facilitated diffusion must
take place in these cases.
Several mechanisms, differing in details, have been proposed.
All of them essentially involve
two steps: the binding to a random non-specific
DNA site and the diffusion (sliding) along the DNA chain.
These two steps may be reiterated
many times before proteins actually find their target, since the sliding is
occasionally interrupted by dissociation.
Berg~\cite{berg81} and Zhou~\cite{zhou04} 
have provided thorough (but somewhat sophisticated) theories
that allow estimates for the resulting reaction rates.
Recently, Halford and Marko have presented
a comprehensive review on this subject and proposed
a remarkably simple and semiquantitative approach
that explicitly contains the mean sliding length
as a parameter of the theory~\cite{halford04}. This 
approach has been refined and put onto a rigorous base
in a recent work by the authors~\cite{klenin05}. 
A plethora of scaling regimes have been studied
for a large range of chain densities and protein-chain
affinities in a recent work by Hu et al.~\cite{hu06}.

The numerical treatment of such a reaction is efficiently
done with the method of excess collisions~\cite{merlitz06}
(MEC), where the reverse process (protein departs from
the binding site and propagates toward the periphery of
the cell) is simulated. This approach delivers exact results
and a significant speed up when compared to straight forward
simulations. Unfortunately, once very large systems are under 
investigation, the numerical treatment of the DNA chain
(whose length is proportional to the volume of 
the cell) quickly turns into a bottleneck, since the MEC
approach requires the construction of the cell in its full
extent. Realistic cell models have to deal with
thermal fluctuations of the chain and its hydrodynamic 
interaction, thereby imposing a strict limit to the size  
that can be managed.
In the present work we demonstrate how to implement
a modification of the MEC approach that allows
to simulate a test system of reasonable size, followed by
an extrapolation to cells of arbitrary size. 

After a definition of the problem in Sect.\ \ref{sec:model},
the MEC approach is briefly summarized in Sect.\ \ref{sec:mec}.
In Sect.\ \ref{sec:facilitated} the numerical implementation
of facilitated diffusion is presented, and \ref{sec:tauf} 
delivers an analytical estimate for the reaction time.
As a preparation for the random walk simulations, the chain
is constructed in Sect.\ \ref{sec:chain} and the specific
recurrence times are evaluated inside a small test system
(Sect.\ \ref{sec:tautilde}). In Sect.\ \ref{sec:diff},
random walk simulations are carried out in order to
construct the empirical MEC-E equations. These are
then employed to extrapolate the reaction times to
cells of much larger dimensions in Sect.\ \ref{sec:results}.
A comparison with exact solutions (in the case of free 
diffusion) and the analytical estimate of Sect.\ \ref{sec:tauf}
suggests that the MEC-E approach delivers an accuracy
of 10 to 15 percent with a speed up of several orders
of magnitude.

\section{Methodology}  \label{sec:theory}
\subsection{Definition of the system} \label{sec:model}
As a {\it cell} we define a spherical volume of radius $R$,
containing a chain ('DNA') of length $L$ and a specific binding 
target of radius $R_a$. The target is located in the
middle of the chain, that in turn coincides with the
center of the cell. The state of the system is well defined
with the position of a random walker ('protein'), which 
can either diffuse freely inside the cell or, temporarily,
associate with the chain to propagate along the chain's contour 
(the numerical realization of this process is discussed in
detail in Sect.\ \ref{sec:facilitated}). The distance of the
walker from the center defines the (radial) reaction 
coordinate $r$. We shall further
denote the periphery of the central target (at  
$r = R_a$) as {\bf state A} and the periphery 
of the cell ($r = R$) as {\bf state B}. To be investigated
is the average reaction time $\tau_{\rm BA}$ the walker
needs to propagate from {\bf B} to {\bf A} as a function
of the binding affinity between walker and chain.

\subsection{Method of excess collisions (MEC)} \label{sec:mec}
The MEC approach was presented in its full generality 
elsewhere~\cite{klenin04,merlitz06}.
In short, it allows to determine the reaction time $\tau_{\rm BA}$ 
while simulating the back reaction {\bf A} $\rightarrow$ {\bf B}
(average reaction time: $\tau_{\rm AB}$) using the relation
\begin{equation} \label{eq:5}
\tau_{\rm BA} = (N_{\rm coll}+1)\cdot \tau_R - \tau_{\rm AB}\;.
\end{equation}
The walker starts at the center ($r(t=0) = 0$) and 
propagates towards the periphery ($r(t=\tau_{\rm AB}) = R$), a process
that is much faster than its reversal ($\tau_{\rm AB} \ll \tau_{\rm BA}$).
On its way to {\bf B}, the walker may repeatedly return back 
to {\bf A}; such an event is called collision, and $N_{\rm coll}$
stands for the average number of collisions. $\tau_R$ is 
the recurrence time and evaluated via  
\begin{equation} \label{eq:10}
\tau_R = \tilde{\tau}_R\, V_{\rm eff}(R)\;,
\end{equation} 
where we have defined the specific recurrence time
\begin{equation} \label{eq:15}
\tilde{\tau}_R \equiv \frac{\tau^*_R}{V_{\rm eff}(R_a)}\;,
\end{equation} 
a quantity, which is derived from simulations of the recurrence
time $\tau^*_R$ within a small test system of the size 
of the central target (Sect.\ \ref{sec:tautilde}). 
The effective volume is defined as
\begin{equation} \label{eq:20}
V_{\rm eff} \equiv \int_{V}
\exp\left[\frac{-U({\bf r})} {k_{\rm B} T}\right]\, d{\bf r}\;,
\end{equation}
and depends upon the energy of the walker $U({\bf r})$ and hence
the implementation of the binding potential between walker 
and chain. 

\subsection{Simple model for facilitated diffusion
of DNA-binding proteins} \label{sec:facilitated}
The nonspecific binding of the walker to the chain is accounted 
for by the attractive step potential
\begin{equation} \label{eq:25}
U(s) = \left\{
\begin{array}{ccl}
-E_o &\hspace{0.8cm} &s \leq r_c \\
0    &\hspace{0.8cm} &s > r_c \;,\\
\end{array}  
\right. 
\end{equation}
where $s$ is the shortest distance between walker and chain.
This defines a pipe with radius $r_c$ around the chain contour 
that the walker is allowed to enter freely from outside, but to exit 
only with the probability    
\begin{equation} \label{eq:30}
p = \exp (-E_o/k_{\rm B} T) \;,
\end{equation}
where $k_{\rm B} T$ is the Boltzmann factor, otherwise it is 
reflected back inside the chain. We
may therefore denote $p$ as {\it exit probability}.
This quantity allows to define the equilibrium
constant $K$ of the two phases, the free and the non-specifically
bound protein, according to 
\begin{equation} \label{eq:35}
K \equiv  \frac{\sigma}{c} = \frac{V_c}{L}\, \left(\frac{1}{p}-1\right)\;,
\end{equation}
where $c$ is the concentration of free proteins and $\sigma$
the linear density of non-specifically bound proteins.
$V_c = \pi\, r_c^2\, L$ is the geometric volume of the chain.
It should be noted that in our previous publication~\cite{merlitz06}, 
$\sigma$ was defined as $\sigma = c\,V_c/(p\,L)$, with the disadvantage
of being non-zero in case of vanishing protein-chain interaction ($p=1$). 
The present choice defines $\sigma$ as the excess concentration of proteins
along the chain contour and leads to a vanishing sliding-length 
(Eq.\ \ref{eq:60}) in case of free diffusion.

The specific binding site is a spherical volume, located in the 
middle of the chain and of identical radius, i.e.\ $R_a = r_c$.
Applying the walker-chain potential Eq.\ (\ref{eq:25}), the effective
volume Eq.\ (\ref{eq:20}) of the cell becomes
\begin{equation} \label{eq:40}
V_{\rm eff}(R) = V + V_c\, \left(\frac{1}{p} - 1 \right)\;,
\end{equation}
and that of the central target is simply
\begin{equation} \label{eq:45}
V_{\rm eff}(R_a) = \frac{V_a}{p} = \frac{4\pi\, R_a^3}{3p}\;.
\end{equation}

\subsection{Analytical estimate for the reaction time and
definition of the sliding length}
\label{sec:tauf}
In case of free diffusion and for a spherical cell, 
Szabo et al.\ have evaluated the exact solution for 
the time a walker needs to reach the radius $R_a$, after 
starting at the periphery $R$, yielding~\cite{szabo80} 
\begin{equation}\label{eq:90}
\tau_{\rm Sz}  =  \frac{R^2}{3\,D}\cdot 
\left(\frac{R}{R_a} + \frac{R_a^2}{2R^2} -\frac{3}{2} \right)\;.
\end{equation} 
Here, $D$ is the diffusion coefficient. In presence of
the chain, exact solutions are known for simple geometrical
setups only~\cite{berg81}, but as discussed elsewhere~\cite{klenin05},
it is still possible to approximate the reaction time
using an analytical approach, once certain
conditions are satisfied. The resulting expression is
\begin{equation} \label{eq:50}
\tau_{\rm BA}(\xi) = \left( \frac{V}{8D_{\rm 3d}\,\xi} +
    \frac{\pi\,L\, \xi}{4 D_{\rm 1d}} \right) \left[
    1 - \frac{2}{\pi} \arctan \left(\frac{R_a}{\xi}\right)\right]\;
\end{equation}
with the 'sliding' variable
\begin{equation} \label{eq:55}
\xi = \sqrt{\frac{D_{\rm 1d}\, K}{2\pi\, D_{\rm 3d}}}\;
\end{equation}
and $D_{\rm 1d}$ and $D_{\rm 3d}$ being the diffusion coefficients
in sliding-mode and free diffusion, respectively. Generally,
the equilibrium constant $K$ has to be determined in simulations
of a (small) test system, containing a piece of chain without specific
binding site. In the present model, $K$ is
known analytically via Eq.\ (\ref{eq:35}). If the 
step-size $dr$ of the random walker is 
equal both inside and outside the chain
(the direction of the step being arbitrary), we further have
\begin{equation} \label{eq:56}
D_{\rm 1d} = D_{\rm 3d} = \frac{dr^2}{6}\;, 
\end{equation}
and hence obtain
\begin{equation} \label{eq:60}
\xi = \sqrt{\frac{r_c^2}{2}\,\left(\frac{1}{p}-1\right)}\;.
\end{equation}
This variable has got the dimension of length; as we have pointed
out in~\cite{klenin05}, it corresponds to the average {\it sliding 
length} of the protein along the DNA contour in the model of Halford 
and Marko~\cite{halford04} and we shall henceforth use the same
expression for $\xi$. In case of free diffusion ($p = 1$), the
sliding length is zero and Eq.\ (\ref{eq:50}) simplifies to
\begin{equation}\label{eq:61}
\tau_{\rm BA}(\xi = 0)  =  \frac{R^3}{3\,R_a\,D_{\rm 3d}}\;,
\end{equation} 
which equals Szabo's result Eq.\ (\ref{eq:90}) in leading order
of $R/R_a$.

\section{Numerical Model}   \label{sec:chain}
In order to approximate the real biological situation,
the DNA was modeled as a chain of straight segments of equal length $l_0$.
Its mechanical stiffness was introduced in terms of a bending energy
associated with each chain joint:
\begin{equation} \label{eq:65}
E_b = k_{\rm B} T \, \alpha \, \theta^2\;,
\end{equation}
where $\alpha$ represents the dimensionless
stiffness parameter, and $\theta$ the bending angle. The numerical
value of $\alpha$ defines the persistence length ($l_p$), 
i.e.\ the ``stiffness'' of the chain.
The excluded volume effect was taken into account by introducing
the effective chain radius $r_c$. The conformations of the chain,
with distances between non-adjacent segments smaller than $2r_c$,
were forbidden.
The target of specific binding was assumed to lie exactly in the
middle of the DNA.
The whole chain was packed in a spherical volume (cell) of radius $R$
in such a way that the target occupied the central position.

\begin{figure}[t]
\centerline{
\includegraphics[width=1.0\columnwidth]{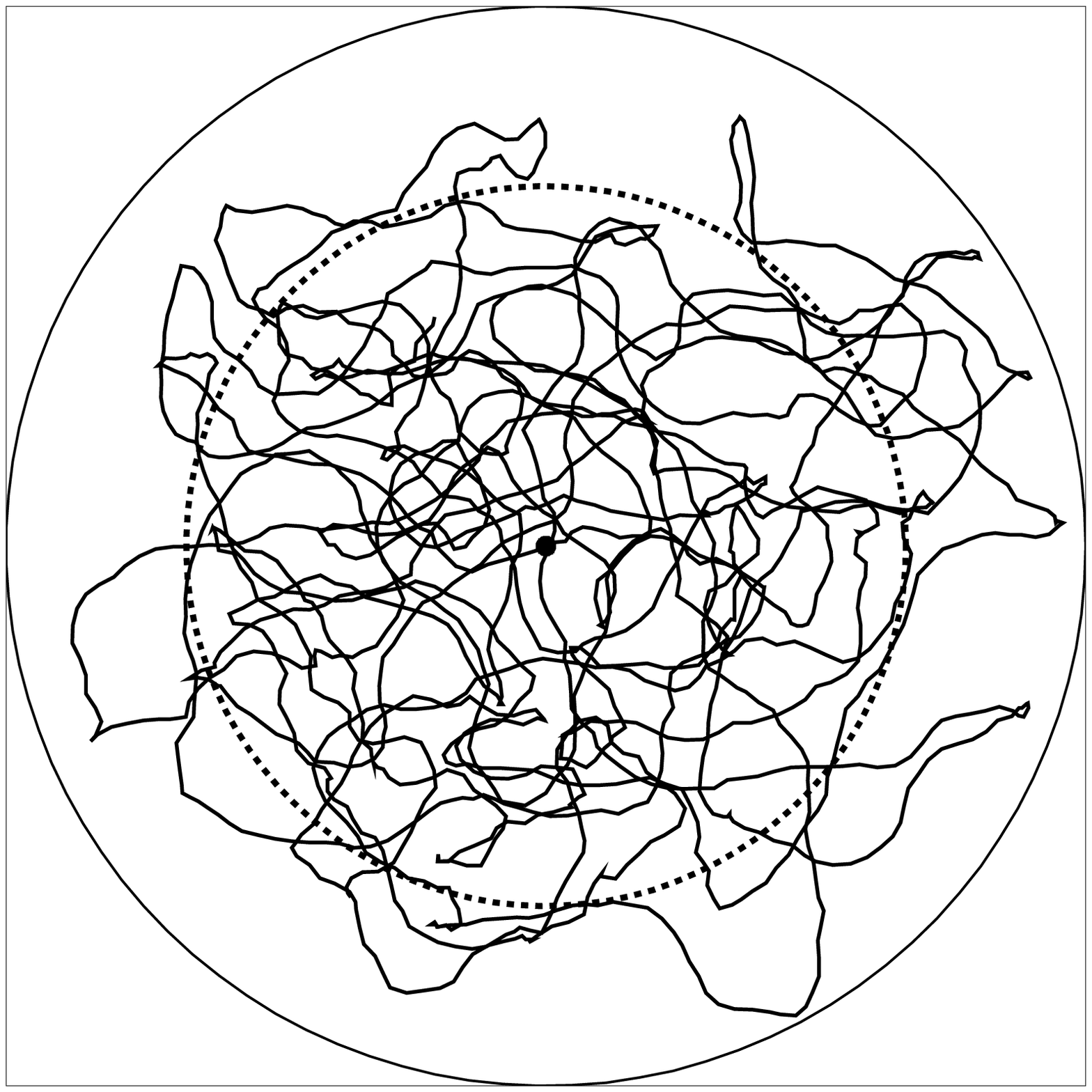}
\vspace{-0.2\columnwidth}
}
\centerline {
\includegraphics[width=1.0\columnwidth]{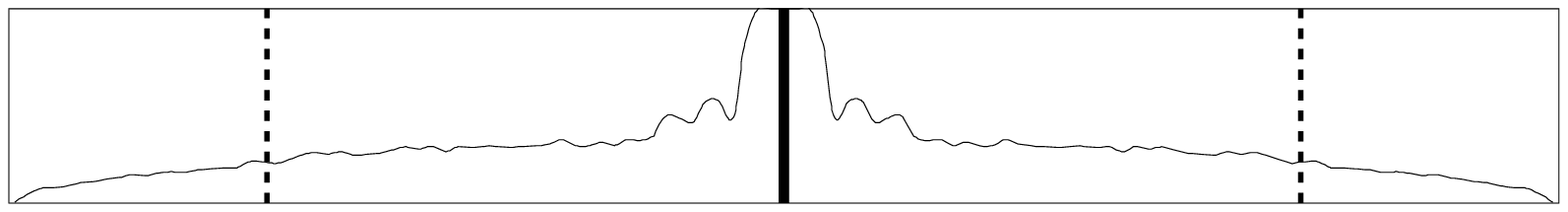}
}
\caption{Upper part: 2-dimensional projection 
of a 3-dimensional random chain-contour of length $L = 400.2$ 
(persistence lengths) confined inside a spherical cell of 
radius $R = 6$. Lower part: Radial chain density distribution,
averaged over 20 conformations. Beyond $r=4$ (dashed line), the
density declines rapidly.
\label{fig:chain}
}
\end{figure}

To achieve a close packing of the chain inside the
cell, we used the following algorithm. First,
a relaxed conformation of the free chain was produced
by the standard Metropolis Monte-Carlo (MC) method.
For the further compression, we defined the
center-norm (c-norm) as the maximum distance from the target
(the middle point) to the other parts of the chain.
Then, the MC procedure was continued with one modification.
Namely, a MC step was rejected if the
c-norm was exceeding 105\% of the lowest value registered so
far. The procedure was stopped when the desired degree of compaction
was obtained.

Below in this paper, one step $dt$ was chosen as the unit of time and
one persistence length $l_p = 50$ nm of the DNA chain as the unit of distance.
The following values of parameters were used. The length of one segment
was chosen as $l_0 = 0.2$, so that one persistence length was partitioned into
5 segments. The corresponding value of the stiffness parameter was
$\alpha = 2.403$~\cite{klenin98}.
The chain radius was $r_c = 0.06$, and the active site
was modeled as a sphere of identical radius $r_a = 0.06$ embedded 
into the chain. The step-size of the random walker both
inside and outside the chain was $dr = 0.02$, corresponding to a diffusion 
coefficient $D_{\rm 3d} = D_{\rm 1d} = dr^2/6 = 2 \cdot 10^{-4}/3$.

Figure \ref{fig:chain} displays a typical chain,
and the radial chain density, obtained with Monte Carlo
integration and averaged over 20 different chain
conformations. The strong increase of chain density towards
the center is merely a geometric effect and caused by 
the chain passing through the origin. Close to the periphery
of the cell, the density was rapidly declining since the contour
was forced to bend back inwards. Within a radius of $r < 4$,
however, the chain content remained reasonably constant, and the 
medium could be regarded as approximately homogeneous.

\section{Computation of the specific recurrence time}
\label{sec:tautilde}
To compute the specific recurrence time $\tilde{\tau}_R$ of
Eq.\ (\ref{eq:15}), the recurrence time inside a small
test system (here: the central binding target of radius $R_a$)
has to be determined. To achieve that, the entire
system, i.e.\ the spherical target and a short piece of chain, was
embedded into a cube of $4 R_a$ side-length with reflective 
walls. In principle, the size of the cube should be of no 
relevance, but it was found that, if chosen too small,
effects of the finite step-size were emerging. The walker
started inside the sphere. Each time upon leaving the spherical
volume a collision was noted. If the walker was about to exit
the cylindrical volume of the chain, it was reflected
back inside with the probability $1-p$ (Eq.\ \ref{eq:30}).
The clock was halted as long as the walker moved outside
the sphere and only counted time-steps inside the sphere.
The resulting recurrence time $\tau^*_R$ has to
be divided by the effective volume of the central target,
Eq.\ (\ref{eq:45}), to yield the specific recurrence time
$\tilde{\tau}_R$. Table \ref{tab:simu} contains the
results for a set of different walker-chain affinities.

\section{Diffusion inside the cell} \label{sec:diff}
The goal is to analyze the propagation of the walker within
a small cell of radius $R_S$ and to extrapolate the results to a
larger system of arbitrary size $R_L > R_S$. As a test site
we have set up a cell of radius $R = 6$, containing a chain
of length $L = 400.2$ (Figure \ref{fig:chain}). The walker
was starting at the center ($r = 0$) and moving towards the
periphery of the cell. Such a process shall be denoted as {\it run}.   
Whenever the walker returned back to the binding site 
($r < R_a$), one collision was noted. A set of 2000 runs,
including 20 different chain conformations, 
was carried out for each value of the exit parameter
$p$, which is related to the walker-chain affinity via
Eq.\ (\ref{eq:30}). For a set of reaction coordinates $r_i$,
the first arrival times were monitored, as well as the
number of collisions that had occurred before first 
passage.

\begin{figure}[t]
\centerline{
\includegraphics[width=1.0\columnwidth]{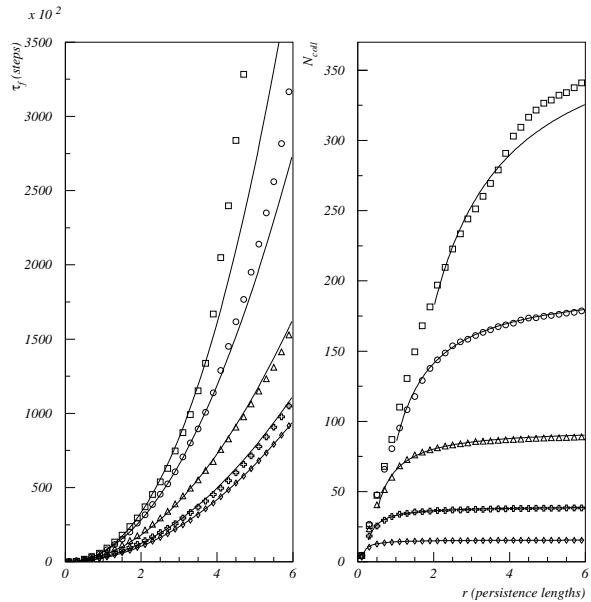}
}
\caption{First passage times (left) and number of collisions
(right) as a function of the reaction coordinate $r$, for
various exit probabilities $p = 2^{-l}$ and $l = $ 3, 5, 7, 9,
11 (bottom to top plots). The curves are $\chi^2$-fits of
Eq.\ (\ref{eq:70}) (left) and Eq.\ (\ref{eq:80}) (right)
within the range $\xi < r < 4$ and extrapolated to $r = 6$.  
\label{fig:fits}
}
\end{figure}

\subsection{The effective diffusion coefficient} \label{sec:Deff}
Figure \ref{fig:fits} displays the first arrival times (left)
for different exit probabilities $p$. To analyse the diffusive
properties of the propagation, the arrival times were fitted
using the macroscopic diffusion law
\begin{equation} \label{eq:70}
\tau_f(p,r) = \frac{r^\alpha}{6D_{\rm eff}(p)}
\end{equation} 
with an effective diffusion coefficient $D_{\rm eff}(p)$. For
low and moderate values of the walker-chain affinity, the
arrival times were well described when assuming regular diffusion,
i.e.\ an exponent of $\alpha = 2$. At high walker-chain
affinities, this exponent was growing larger, indicating the
onset of anomalous subdiffusion. Table \ref{tab:simu} contains
the fit parameters when the fits were carried out within the
range $\xi < r < 4$, and the solid curves in figure 
\ref{fig:fits} (left) display the resulting functional form
of Eq.\ (\ref{eq:70}), when extrapolated to the full range
$0 < r < 6$.  

The lower boundary of the fit range, the sliding length $\xi$,
was implemented because the near the central
target, the transport process was dominated 
by one dimensional sliding rather than three dimensional diffusion.
The upper boundary was introduced since the chain distribution
beyond $r > 4$ was affected by boundary effects near the periphery 
of the cell, as is clearly visible at Figure \ref{fig:chain}. 
Within the range of $\xi < r < 4$, however, 
the propagation of the walker could approximately
be regarded as a random walk inside a homogeneous and 
crowded environment.

\begin{table}[b]
\caption{
The first column is the exponent of the exit probability $p = 2^{-l}$,
the second column the corresponding sliding parameter,  
followed by the specific recurrence time
(Sect.\ \ref{sec:tautilde}). The next six columns contain
optimized parameters of the $\chi^2$-fits of equations 
(\ref{eq:70}), (\ref{eq:80}) and (\ref{eq:81}). The last
column defines the speed up achieved with the extrapolation
from $R_S = 4$ to $R_L = 6$, when compared with the explicit
simulation of the reaction time $\tau_{\rm BA}(R_L)$.
\label{tab:simu}}
\begin{center}
\begin{tabular}{r|cc|cc|cc|cc|c}
$l$ & $\xi $& $\tilde{\tau}_R$ & $D_{\rm eff}^{1)}$ & 
$\alpha$ & $R_{\rm eff}$& 
$N_{\infty}$ & $D_{\rm eff}^{1)}$ &  $R_{\rm eff}$ & 
$\frac{\tau_{\rm BA}(R_L)}{\tau_{\rm AB}(R_S)}$
\\ \hline
Eq.&(\ref{eq:60})&(\ref{eq:15})
  &\multicolumn{2}{c|}{(\ref{eq:70})}  
  &\multicolumn{2}{c|}{(\ref{eq:80})} 
  &\multicolumn{2}{c|}{(\ref{eq:81})} &  \\ \hline
0 & 0     & 4464 & 6.63 & 2   &$0.064$ & $3.83$ & 6.09 &0.060&520  \\
1 & 0.042 & 2594 & 6.66 & 2   &$0.078$ & $6.42$ & 5.98 &0.069&410  \\
2 & 0.073 & 1413 & 6.55 & 2   &$0.084$ & $9.95$ & 6.42 &0.079&354  \\
3 & 0.112 & 741.6& 6.35 & 2   &$0.100$ & $15.7$ & 6.73 &0.092&292 \\
4 & 0.164 & 379.7& 5.97 & 2   &$0.118$ & $25.0$ & 6.42 &0.116&221 \\
5 & 0.236 & 192.6& 5.37 & 2   &$0.174$ & $39.7$ & 5.27 &0.167&169 \\
6 & 0.337 & 96.81& 4.50 & 2   &$0.224$ & $61.9$ & 4.28 &0.231&120 \\
7 & 0.478 & 48.62& 3.67 & 2   &$0.319$& $95.3$  & 2.95 &0.348&94 \\
8 & 0.677 & 24.30& 2.83 & 2   &$0.417$& $135$   & 2.09 &0.491 &70 \\
9 & 0.959 & 12.17& 2.47 & 2.07& $0.59$& $199$   & 1.28 &0.69&61 \\
10& 1.357 & 6.089& 2.44 & 2.20& $0.81$& $279$   & 0.63 &1.10&54 \\
11& 1.920 & 3.044& 2.41 & 2.27& $1.10$& $398$   & 0.35 &1.50&63 \\

\end{tabular}
\end{center}
\flushleft 1) In units of $10^{-5}$
\end{table}

\subsection{The functional dependence of $N_{\rm coll}$ on the 
target-distance}
\label{sec:Ncoll}
The right hand side of Figure \ref{fig:fits} displays the 
number of collisions $N{\rm coll}$ as a function of the radius $r$
for various walker-chain affinities. Quite generally, there exists
a steep increase close to the central target, after
which the function gradually levels off to reach a plateau. 
In Appendix \ref{sec:a1}, we argue that this functional behavior
can be described as 
\begin{equation} \label{eq:80}
N_{\rm coll}(r) = \frac{N_{\infty}\cdot (r - R_{\rm eff})}{r}\;,
\end{equation} 
where $N_{\infty}$ stands for the asymptotic limit 
$N_{\rm coll}(r\rightarrow \infty)$ and
$R_{\rm eff}$ defines an effective target size.
As a result of facilitated diffusion, the mode of propagation
is predominantly one-dimensional near the central target.
This relation is therefore invalid within a radius of the
average sliding length of the walker and should be
applied for $r > \xi$. Under this condition, both $N_{\infty}$
and $R_{\rm eff}$ were used as free fit-parameters and
the fit range was restricted to $\xi < r < 4$, for the same
reason as discussed in Sec.\ \ref{sec:Deff}. The solid
curves of Figure \ref{fig:fits} (right) display the best fits
(extrapolated to $r = 6$), and Table \ref{tab:simu} contains
the corresponding values for the fit-parameters.

An alternative approach to $N_{\rm coll}(r)$ is described in
Appendix \ref{sec:a2}, leading to 
\begin{equation} \label{eq:81}
N_{\rm coll}(r) = \frac{f(r)}{V_{\rm eff}\, \tilde{\tau}_R}\;,
\end{equation} 
where $f(r)$ is defined in Eq.\ (\ref{eq:a60}). It
contains both parameters $D_{\rm eff}$ 
and $R_{\rm eff}$ which are
used as free fit parameters to determine the effective diffusion
coefficient and an effective target size.
The results are given in Table \ref{tab:simu}. The effective
volume $V_{\rm eff}(r)$ as a function of radius $r$ is actually a  
complicated function that depends on the radial chain density 
(Fig.\ \ref{fig:chain}), but for this investigation we have 
assumed a perfectly homogeneous chain density and evaluated
\begin{equation} \label{eq:82}
V_{\rm eff}(r) = \frac{V(r)\, V_{\rm eff}(R)}{V(R)}
\end{equation}
with the cell-radius $R=6$.
When comparing the best fits for the effective diffusion
coefficient with the results of Eq.\ (\ref{eq:70}), the
agreement is only qualitative. In fact, Eq.\ (\ref{eq:82})
does not deliver an accurate way to determine $D_{\rm eff}$.
This may be so because the second term of function
\begin{equation} \nonumber
f(r) = \frac{r^2}{3\,D_{\rm eff}}\cdot 
\left(\frac{r}{R_{\rm eff}} + \frac{R_{\rm eff}^2}{2\,r^2} - 1 \right)\;,
\end{equation}
the fraction $R_{\rm eff}^2/r^2$, quickly drops to zero and hence
both fit-parameters $D_{\rm eff}$ and $R_{\rm eff}$ become linear 
dependent. This implies that $D_{\rm eff}$ is actually 
determined locally, close to the (effective) target, and not
averaged over $\xi < r < 4$.
Except for high walker-chain affinities, Eq.\ (\ref{eq:70}) 
delivers a more accurate description
of the diffusion process, which is verified with the quadratic 
dependence of the passage time on the reaction coordinate.

The effective target size $R_{\rm eff}$ agrees 
fairly well with the corresponding findings of Eq.\ (\ref{eq:80})
and increases substantially with the walker-chain affinity.
As a consequence of facilitated diffusion, the walker 
initially moves away from the target in one-dimensional
sliding mode, and its (effectively) free diffusion begins
further outside, thereby increasing the effective
target size. Hence it is no surprise to find $R_{\rm eff}$ 
being of similar dimension as the average sliding 
length $\xi$ (Table \ref{tab:simu}).

\subsection{The empirical MEC-E equations} \label{sec:mec-e}
It is now possible to combine Equations (\ref{eq:80}) and 
(\ref{eq:70}) with (\ref{eq:5}) and (\ref{eq:10}) to 
obtain the empirical MEC-E equations
\begin{equation} \label{eq:85}
\tau_{\rm BA}(p,r) = (N_{\rm coll}(p,r)+1)\, 
\tilde{\tau}_R\, V_{\rm eff} - \tau_f(p,r)\;,
\end{equation}
which allow to evaluate the reaction time $\tau_{\rm BA}(p,r)$
for any reaction coordinate $r$ by extrapolation of the
number of collisions $N_{\rm coll}(p,r)$ and the first
arrival times $\tau_f(p,r)$. 

When using Eq.\ (\ref{eq:81})
instead of (\ref{eq:80}), we obtain
\begin{equation}\label{eq:86}
\tau_{\rm Sz,eff}(p,r)  =  \frac{r^2}{3\,D_{\rm eff}(p)}\cdot 
\left(\frac{r}{R_{\rm eff}(p)} + \frac{R_{\rm eff}^2(p)}
{2r^2} -\frac{3}{2} \right)\;,
\end{equation} 
which can be regarded as an empirical generalization 
of Szabo's exact result for free diffusion, Eq.\ (\ref{eq:90}). 

Since both sets of equations are based on the MEC approach, 
while employing different ways to extrapolate the number 
of collisions to large cells, we will refer to them
as MEC-E equations.
In the following section we will apply both approaches,
Eq.\ (\ref{eq:85}) and Eq.\ (\ref{eq:86}), to extrapolate
the reaction times to large cell radii, and compare
their results.

\section{Results} \label{sec:results}

\begin{figure}[t]
\centerline{
\includegraphics[width=1.0\columnwidth]{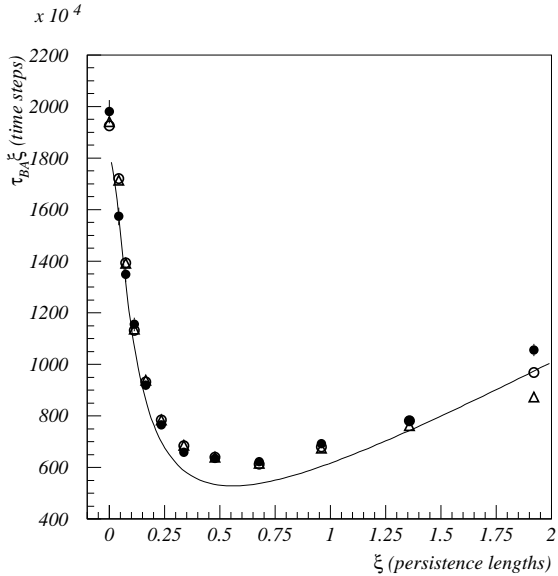}
}
\caption{Reaction time $\tau_{\rm BA}$ of the protein as a
function of the sliding length Eq.\ (\ref{eq:60}). The explicit
simulation (solid dots) required about 140 times the number of 
simulation steps of the extrapolation using
Eq.\ (\ref{eq:85}) (open circles) or Eq.\ (\ref{eq:86}) (triangles).
The curve is the analytical estimate Eq.\ (\ref{eq:50}).
\label{fig:tauba}
}
\end{figure}

As a first consistency-check, the MEC-E equations
were applied to estimate the reaction time $\tau_{\rm BA}$ of
the walker entering the cell at radius $R = 6$. The simulation
of the reaction {\bf B} $\rightarrow$ {\bf A}
was additionally carried out explicitly and the results are displayed
in Figure \ref{fig:tauba}. The
effective volume of the cell was evaluated via Eq.\ 
(\ref{eq:40}), using the total chain length $L = 400.2$.
The results, shown in  Figure \ref{fig:tauba}, imply that 
the extrapolation from $R_S = 4$ (the
radius used to optimize the parameters) to $R_L = 6$
delivered accurate results for the reaction times. This should
not be taken for granted, taking into account the problematic chain
distribution between $R_S < r < R_L$. In fact, $\tau_f(p,r)$
becomes inaccurate in this region due to anomalous diffusion
(Figure \ref{fig:fits}, left), but this term contributes just
a small amount to Eq.\ (\ref{eq:85}), since for 
reasonably large cells the first arrival time  $\tau_f$
is small compared to the corresponding reaction time 
$\tau_{\rm BA}$. Its error was therefore of little impact.
On the other side, the collisions $N_{\rm coll}(r)$ with 
the central target, which form the main contribution to
Eq.\ (\ref{eq:85}), 
were much less affected by the chain distribution far outside 
the center (Figure \ref{fig:fits}, right)
and were extrapolated accurately, despite of 
the sparse chain density at the cell periphery.
This feature contributes to the fact that the extrapolation
process appears to be insensitive to the chain distribution 
far away from the target. Similarly, Eq.\ (\ref{eq:86}) 
delivered consistent and accurate results, except for the
last data point which belongs to the highest walker-chain
affinity. Here, the sliding-length already reaches one half
of the system size that was used to fit the empirical 
parameters. A larger dimensioned test system is required 
for such high affinities to 
increase the accuracy of the extrapolation procedure. 

The simulation
time required to set up the MEC-E equations (\ref{eq:85}) 
and (\ref{eq:86}) equals the
average number of time steps the walker needed to reach the
radius $R_S = 4$ when starting at the central target, 
i.e. $\tau_f(p,R_S)$. Compared to the corresponding 
time required to simulate $\tau_{\rm BA}(R_L)$ explicitly,
a speed up between $50$ and $500$ was gained, depending 
upon walker-chain affinity (Table \ref{tab:simu}, last column). 
Integrated over all 12 data points, a total speed up of $140$ was
derived.  

\begin{figure}[t]
\centerline{
\includegraphics[width=1.1\columnwidth]{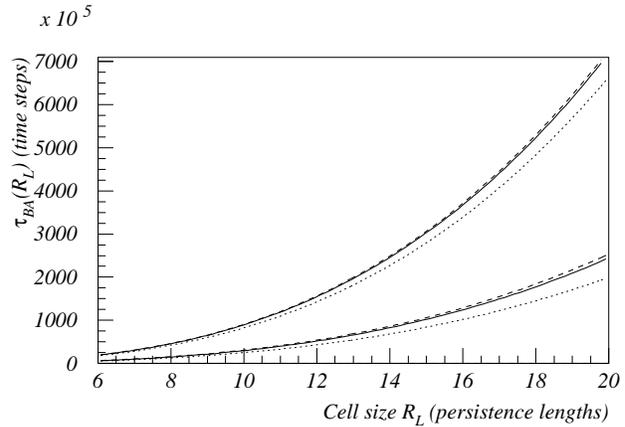}
}
\caption{Extrapolation of the reaction time $\tau_{\rm BA}$ to
large cell radii $R_L$. The dotted curve is the analytical estimate 
Eq.\ (\ref{eq:50}), the solid and dashed curves correspond to 
the MEC-E equations
(\ref{eq:85}) and (\ref{eq:86}), respectively. 
Upper triple: $p=1$ (free diffusion). Lower triple: $p = 2^{-8}$, 
where facilitated diffusion is most effective.
\label{fig:tauflex}
}
\end{figure}

It is possible and intended to exploit this method for
extrapolations to much larger systems. Figure 
\ref{fig:tauflex} displays
the extrapolation of $\tau_{\rm BA}(p,R_L)$ up to
$R_L = 20$ for $p=1$ (free
diffusion) and $p=2^{-8}$, close to the minimum in
Figure \ref{fig:tauba}. The chain density was assumed 
to remain constant,
i.e.\ its length was growing as $L(R_L) \sim R_L^3$.
Explicit simulations of $\tau_{\rm BA}$ 
are not feasible any more for such large cells.
However, for free diffusion, Eq.\ (\ref{eq:90})
is available, and both extrapolation methods delivered
reaction times about 8\% above the exact solution,
which, in this plot, was un-distinguishable from the
approximation Eq.\ (\ref{eq:61}). When protein-chain
interaction was enabled, both extrapolation methods
delivered almost identical results, which were about
15\% above the analytical estimate Eq.\ (\ref{eq:50}).

\section{Summary}
In this work, the empirical MEC-E equations (\ref{eq:85}) 
and (\ref{eq:86}) were
derived and tested against random walk simulations. 
Whereas the original MEC approach
(Sect.\ \ref{sec:mec}) represents an exact method to obtain the
average reaction time $\tau_{BA}$ by simulating the much faster
back-reaction {\bf A} $\rightarrow $ {\bf B}, it still requires
to set up a model system of full size $R$. This would become
prohibitive in simulations of large cells containing
realistic chains with
thermal fluctuations and hydrodynamic interactions.

We have demonstrated that the simulation of
a test system of moderate size is sufficient to extract
reaction times of much larger cells. This is so because
the number of collisions as a function of the cell radius,
$N_{\rm coll}(r)$, is asymptotically approaching a plateau
(Figure \ref{fig:fits}, right).
In this region, the reaction time is merely proportional to
the effective volume $V_{\rm eff}$, as shown in Eq.\ (\ref{eq:85}),
with a small correction in form of the first passage time
$\tau_f(r)$, Eq.\ (\ref{eq:70}). This quantity is
easily estimated once the effective diffusion coefficient
is determined. If the test system is too small for
$N_{\rm coll}(r)$ to reach the plateau, it is still
possible to obtain accurate results, because the functional
form of this quantity is known (Eq.\ \ref{eq:80} and
\ref{eq:81}), so that extrapolations to
larger cells become feasible. 

The size of the test system has to be chosen with care,
because only those regions are of use in which the walker 
experiences a randomized and approximately homogeneous 
environment. Within the central region, typically of the size 
of the sliding length $\xi$,
the reaction time is dominated by 1-dimensional (sliding) 
instead of 3-dimensional diffusion. This part of the
cell has to be excluded when the walker's diffusion 
properties are analyzed. The same holds true for the 
outermost region, where the chain conformation
exhibits boundary effects. Assuming that the sliding 
length $\xi$ does not exceed the persistence length $l_p$,
a cell radius $R$ of five persistence lengths 
appears adequate. Here, the region $\xi < r < R-2\,l_p$
may be exploited to set up the empirical equations
(\ref{eq:85}) or (\ref{eq:86}). With increasing 
walker-chain affinity and sliding length, the radius $R$
has to be adjusted accordingly.

The results presented above demonstrate how the MEC-E
approach delivers a speed up between 50 and 500 (depending on
walker-chain affinity, Table \ref{tab:simu}) by extrapolation
from $R_S=4$ to $R_L=6$, with respect to explicit simulations
of the reaction time $\tau_{\rm BA}$. With increasing 
radius $R_L$, Eq.\ (\ref{eq:86}) is approximated as
\begin{equation}\label{eq:95}
\tau_{\rm Sz,eff}(R_L\gg R_{\rm eff}) \approx  
\frac{R_L^3}{3\,D_{\rm eff}\, R_{\rm eff}} \;, 
\end{equation}  
and the speed up is therefore approximately growing
proportional to $R_L^3$.

\appendix
\section{Proof of equation (\ref{eq:80})} \label{sec:a1}
As was shown by Berg~\cite{berg93},  the
probability of a walker, after starting at $r_{\rm ini}$
(where $R_a < r_{\rm ini} < R$), to be adsorbed at
$R_a$, before reaching the distance $R$, is
\begin{equation} \label{eq:a05}
P(R) = \frac{R_a(R - r_{\rm ini})}{r_{\rm ini}(R-R_a)}\;.
\end{equation}
This was derived from
the steady-state solution of Fick's second 
equation for spherical symmetry,
\begin{equation} \label{eq:a10}
\frac{1}{r^2}\, \frac{d}{dr}\;\left(r^2\, \frac{dC(r)}{dr}\right) = 0\;.
\end{equation} 
Here, $C(r)$ is the concentration,
having a maximum at the particle source radius $r=r_{\rm ini}$ and
dropping to zero at the adsorbers radii $r = R_a$ and $r = R$.

In our case, not the probability
$P(r)$, but the average number 
$N_{\rm coll}(r)$ of events in which
the walker returns to $r = R_a$ before first reaching
the distance $r = R$ is of interest. We shall now assume
that $N_{\rm coll}(r)$ is known for one particular distance
$r$, and we want to derive $N_{\rm coll}(r+dr)$. The probability,
that the walker, starting from $r$, goes straight to $r+dr$,
is $1-dP(r)$. Then, the probability to first return back to
the target, before passing through $r$ and reaching $r+dr$, is
$dP\, (1-dP)$. In this latter case, $2\cdot N_{\rm coll}(r) +1$
collisions have already occurred in average. The probability to return
exactly $n$ times to the target and back to $r$ before 
reaching $r+dr$ is $dP^n\, (1-dP)$,
yielding $(n+1)\cdot N_{\rm coll}(r) + n$ collisions. The sum
\begin{eqnarray} \nonumber
N_{\rm coll}(r+dr) &=& (1-dP)\sum_{n=1}^{\infty} [n\,N_{\rm coll}(r) + n-1]\, 
dP^{n-1} \\
&=& \frac{N_{\rm coll}(r)+1}{1-dP} - 1
 \label{eq:a15} 
\end{eqnarray}  
leads to the differential equation
\begin{equation} \label{eq:a20}
\frac{dN(r)}{dr} = N(r)\, \frac{dP}{dr} + \frac{dP}{dr}\;.
\end{equation}
With Eq.\ (\ref{eq:a05}) we further have
\begin{equation} \label{eq:a25}
dP = \frac{R_a\, dr}{r\, (r - R_a)}\;,
\end{equation} 
so that Eq.\ (\ref{eq:a20}) is solved as
\begin{equation} \label{eq:a30}
N_{\rm coll}(r) = (N_\infty + 1)\,  \frac{(r - R_a)}{r} -1\;.
\end{equation}
Here, $N_\infty = N_{\rm coll}(r\rightarrow \infty)$ is the asymptotic
limit for the number of collisions far away from the target.
This solution is incorrect close to the target, where 
$N_{\rm coll}(R_a) = -1$. In fact, the validity of this  approach 
is restricted to length scales that are large compared to the 
(finite) step-size. In particular,  
since we want extrapolate $N_{\rm coll}(r)$ to
a large distance, we can assume $r$ to be large enough so that
$N_{\rm coll}(r) \gg 1$. Then, the sum Eq.\ (\ref{eq:a15}) 
simplifies to
\begin{eqnarray} \nonumber
N_{\rm coll}(r+dr) &=& (1-dP)\sum_{n=1}^{\infty} n\,N_{\rm coll}(r)\, 
dP^{n-1} \\
&=& \frac{N_{\rm coll}(r)}{1-dP}\;,
 \label{eq:a35} 
\end{eqnarray}  
leading to 
\begin{equation} \label{eq:a40}
\frac{dN(r)}{dr} = N(r)\, \frac{dP}{dr}\;,
\end{equation}
which finally solves to
\begin{equation} \label{eq:a45}
N_{\rm coll}(r) = N_\infty\; \frac{(r - R_a)}{r}\;.
\end{equation}
Both parameters $N_\infty$ and $R_a$ were used as free 
fit parameters. We have verified that Eq.\ (\ref{eq:a30}) and
Eq.\ (\ref{eq:a45}) deliver identical results when extrapolating
to large radii, so that, for sake of simplicity, Eq.\ (\ref{eq:a45})
was applied throughout this work.

\section{Proof of equation (\ref{eq:81})} \label{sec:a2}
When considering Eq.\ (\ref{eq:5}), 
\begin{equation} \label{eq:a50} \nonumber
\tau_{\rm BA} + \tau_{\rm AB} = (N_{\rm coll}+1)\cdot \tau_R\;,
\end{equation}
we note that in case of free diffusion the reaction time $\tau_{\rm BA}$
is given by Eq.\ (\ref{eq:90}) and $\tau_{\rm AB}$ by Eq.\ (\ref{eq:70})
with the free diffusion coefficient $D$, Eq.\ (\ref{eq:56}), so that
\begin{equation} \label{eq:a55}
(N_{\rm coll}(r)+1)\cdot \tau_R(r) = f(r) 
\end{equation}
and
\begin{equation} \label{eq:a60}
f(r) = \frac{r^2}{3\,D}\cdot 
\left(\frac{r}{R_a} + \frac{R_a^2}{2\,r^2} - 1 \right)
\end{equation}
with $r > R_a$. Using Eq.\ (\ref{eq:10}) we obtain
\begin{equation} \label{eq:a65}
N_{\rm coll}(r) = \frac{f(r)}{\tilde{\tau}_R\, 
V_{\rm eff}(r)} - 1 \;.
\end{equation}
Both quantities $D$ and the effective source radius $R_a$ are 
used as free fit parameters.

\end{document}